# Autocorrelation Function for Radio Galaxies

W.Godłowski,
Astronomical Observatory Jagiellonian University
ul Orla 171  Kraków Poland
email: godlows@oa.uj.edu.pl

**Abstract.** We discuss the autocorrelation function $\xi(r)$ for the sample 1157 radio-identified galaxies (Machalski & Condon 1999)[1]. The sample of galaxies is based on the Las Campanas Redshift Survey (LCRS)[2] and *1.4 Ghz* NRA)-VLA Sky Survey (NVSS)[3] catalog of the radio sources. For separation *2-15 $h^{-1}$ Mpc* autocorrelation function $\xi(s)$ can be approximated by the power law with correlation length and slope $-\gamma=-1.8$. There are no clearly differences for correlation length between AGN and SB galaxies while value $\gamma$ for both samples are different: $\gamma=2.4$ for AGN and $\gamma=1.7$ for SB.

## 1. Introduction

Studying of the distribution of the radio galaxies is very important problem and such investigations have a long history. Analysis of the spatial clustering gives us important knowledge of the large scale structures in the Universe. Early *2D* study of clustering based on surveys such as 4C and Parkes, gave the first convincing evidence for the large-scale homogeneity of the Universe (Webster 1977)[4]. However, in the *2D* analysis, projection effects are very important and any possible clustering effects are hugely diluted. Clustering of the radio galaxies was detected for the first time by Peacock and Nicholson (1991) (hereafter PN91)[5] in *3D* analysis. They used a redshift survey of 329 galaxies with *z<0.1* and *S(1.4Ghz)>500mJy*. The obtained result was that the correlation function measured in the redshift space has the form $\xi(s)=[s/11h^{-1}Mpc]^{-1.8}$ (where $h \equiv H_0/100$ *km/sec\*Mpc*). This result corresponds to a clustering amplitude being between normal galaxies and rich clusters of galaxies. This agrees with the fact that radiogalaxies are normally found in rich groups of galaxies (Allington-Smith et al 1993)[6]. However, one should note that number of galaxies in the sample of the Peacock and Nicholson was small.

In the 1997 Peacock analyzed a sample of 451 radio identified galaxies selected of the base LCRS (Las Campanas Redshift Survey (Shectman et al. 1996)[2] and NVSS survey (Condon et al. 1998)[3]. The LCRS contains over 90000 galaxies to a limiting isophotal magnitude of 18.2-18.5 of which over 26000 have spectroscopic redshift measured. The LCRS contains six slices approximately 1.5⁰x 80⁰ of which four are sufficiently high declination to overlap with the NVSS survey. Based on the projected correlation function $\Xi(r)=\int \xi[(r^2+x^2)^{0.5}]dx$ he found that distribution of the $\xi(r)$ is in agreement with power law model with correlation length: *5$h^{-1}$Mpc* and slope $-\gamma=-1.8$. The correlation length for radio-loud subsample is approximately *6.5$h^{-1}$MPc*. It means that the radio-loud subset of the LCRS is clustered only slightly more strongly than the whole LCRS which is in apparent conflict with PN91. Peacock (Peacock 1997)[5] suggested that these differences are probably because his sample is dominated by SB galaxies while PN91 sample are mostly luminous AGN.

The aim of our investigation was to analyse possible clustering of the radio galaxies on the base of a sample of 1157 galaxies (Machalski and Condon 1999)[1]. This sample was selected from LCRS using NVSS. Taking into account a strongly increasing incompleteness of the LCRS sample with increasing isophotal magnitude of galaxies, the radio detection was limited to galaxies with $m_{iso} \leq 18.0$ mag. The optical and radio data, supported by the IRAS far-infrared (FIR) data, were then used to classify 1157 detected galaxies by dominant radio-energy source: starburst or AGN. It allowed us to test any possible difference in clustering SB and AGN galaxies.

## 2. Correlation Function

The correlation function measures how clustered the sources are, compared to a random Poisson distribution (Peebles 1980)[7]. The angular two-point correlation function $w(\theta)$ gives the excess probability of finding two sources in the solid angles $\delta\Omega_1$ and $\delta\Omega_2$ separated by an angle $\theta$. Similarly the *3D correlation function* $\xi(r)$ informs us of the influence of a galaxy's presence in the $\delta V_1$ on the probable presence of a galaxy in $\delta V_2$. In practice, correlation function is usually measured by creating a random catalogue, and counting pairs either within catalogues or between two catalogues. There is a lot of methods of the estimated autocorrelation functions $w(\theta)$ and $\xi(r)$. The simplest way to estimate this function are*:* $w(\theta)=(DD/RR)-1$ and $\xi(r)=(DD/RR)-1$ where $DD$ and $RR$ are numbers of data-data pairs separated by distance $\theta+\delta\theta$ or $r+\delta r$ with real and random generated catalogue respectively. Random catalogue is required to have the same selection effects that original one.

Another estimators for autocorrelation function are: $w(\theta)=DD/DR-1$ and $\xi(r)=DD/DR$ where DR is the number of pairs between catalogues. Both methods give the similar results (PN91). Because the second method requires very carefully reproduced complicated shape of the LCRS the first one is used during our work. The number of "objects" in the random catalogue may be much greater than that is in the real catalogues. It allows reducing the errors in the estimates. However, in such cases the above equations are modified:

$$w(\theta)=(DD/RR)(n_r(n_r-1)/n(n-1))-1, \qquad (1)$$

$$\xi(r)=(DD/RR)(n_r(n_r-1)/n(n-1))-1,$$

where $n$ and $n_r$ are average number density in the real and random-generated catalogue respectively. Relations between $w(\theta)$ and $\xi(r)$ are usually complicated and depend on details of the galaxy luminosity function. Also, the structure in three dimensions tends to be smoothed out when projected onto two dimensions. In the special case when $\xi(r) \propto r^{-\gamma}$ is a pure power law on all scales, then $w(\theta) \propto \theta^{-\gamma+1}$ is also a power law which exponent is increased by unity, although its amplitude depends on an integral of the luminosity function (Totsuji & Kihara 1969) [8]. There is also a simple relation (Peebles 1973, 1980) [7,9] for the amplitudes of $w(\theta)$ in statistically homogeneous catalogues observed to different magnitude limits, and therefore to different depths $D$

$$w(\theta)=w(\theta D)/D. \qquad (2)$$

Please note that the knowledge of the numerical value of $\xi(r)$ function without any information about errors is not sufficient. It means that having made estimates of $\xi(r)$ we now need to consider the error bars. In absence of clustering we obtain $<\xi>=0$ and $<\xi^2>=1/N$ where $N$ is the number of independent pairs in a given radial bin (Peebles 1980)[7]. For a non-zero value of autocorrelation function $\xi$, it is suggested the usual 'Poisson error bar'

$$\Delta\xi/(1+\xi)=1/\sqrt{N_p} \qquad (3)$$

This will usually be a lower limit for uncertainty in function $\xi$. Peebles (1973)[9] showed that the right-hand side should be increased by a factor of approximately $1+4\pi nJ_3$, where $4\pi nJ_3$ is the volume integral of $\xi$ out to radius of interest. The problem with this expression is that $J_3$ may be hard to estimate. In our investigation we adopted a value of $J_3$ found for LCRS by Lin et al (1996) [10].During analysis of the correlation function we should take into account the problem of the weighting schemes (Davis and Peebles 1983)[11]. Specifically, the sum of all pairs in a given interval $\Delta r$; DD is given by formulae:

$$DD=\Sigma\Sigma w_i w_j n_i n_j, \qquad (4)$$

where the $n_i$ and $n_j$ are $\delta$ function giving the position of each particle, the sum over $i$ sums over all particle in the sample and the sum over $j$ includes only particle in the proper interval $\delta r$ from particle *i*. Usually, during estimating of the correlation function, two major weighting schemes are used. In the first scheme we count each pairs with the equal weight. In the second scheme we weight each pair *ij* by $\phi_i^{-1}\phi_j^{-1}$, so each part of the space is weighted in proportion to the number of pairs that would be presented in totally volume-limited sample. The second scheme is close to the minimum variance weight for determination of *ξ(r)* (Davis and Huchra 1982) [12]. In the cases of the second scheme, weight should be modified by the factor connected with the effect of the field-to-field sampling fraction and the incompleteness as a function of the apparent magnitude and central surface brightest (Lin et all 1996) [10]. In the cases of the LCRS Bharadwaj, Gupta and Seshadri (1999) [13] proposed original modification by taking into account thickness of the catalogue. They constructed *2D* distribution by collapsing the thickness of the slice and modified weight by the factor 1/*z*. We analyzed different weight scheme and found no significant differences in the results. So for the simplicity we decide to concentrate on the first scheme where every pair has the equal weight.

## 3. Results

### 3a. 2 Dimensional analysis

At first we decided to test our procedure, computing angular correlation function for the sample of the optical galaxies taking from LCRS the slice -12. The result is presented in the fig.1 and fig.1a. We can observe that the power law exponent -*ε*= *-0.67*. It means that if we assume that correlation function *ξ(r)* ∝ $r^{-\gamma}$ is a pure power law on all scales, the -*ε* value is equal -*γ*+*1*. In such a case -*γ*+1= -0.67 what means that the value *γ=1.67* is not far from value 1.52 found by Tucker et al. (1997) [14] for LCRS from *3D* analysis. We also could see that the *w(θ)* breaks from power law on a distance 0.0095rad. According to Bhardawaj, Gupta and Seshadri (1999) [13], the survey extends to the redshift ~0.2 corresponding to 600h$^{-1}$Mpc in the radial direction. This result gives, according to Peebles (1973, 1980), scale of the correlation length *w(θD)* 5.7 h$^{-1}$*Mpc*. One should remember that, when we test the sample of all galaxies in the LCRS, survey extends to ~0.26, instead of ~0.2. It gives the scale of the correlation length *w(θD)* 7.4 h$^{-1}$*Mpc*. It is close to the correlation length found by Tucker et al. (1997) [14] - *6.28 h$^{-1}$Mpc*. Because even from *2D* analysis we obtain result similar to Tucker et al. (1997) [14], it is proved that our procedure is correct. Moreover it shows that even from *2D* correlation function we could obtain correct and important information. It is very important because when we analysed radiogalaxies, the sample is small and comparing results of *2D* and *3D* analysis is crucial.

Now we repeat analysis for our sample of the 1157 radio-identified galaxies. Result of the analysis is presented in the fig.2. We fit power law exponent, obtaining –*ε*=-*γ*+*1* equal -0.97 for all radio galaxies, -1.13 for AGN and -0.86 for SB. The *w(θ)* breaks from the power law on a distance 0.0065*rad* that is significantly less then value obtained for "all" LCRS galaxies. Also it should be noticed that when we take into account only radiogalaxies with known redshifts, function *w(θ)* breaks from the power law at the distance 0.0015*rad* what is too small to obtain correctly the power law exponent. It suggests strong *2D* selection effect during taking out LCRS galaxies for redshift measurement. One should note that using second weighting scheme does not change significantly this result.

### 3b. 3D analysis

Also in this case we decided to start our analysis with optical galaxies. We chose for analysis the LCRS -12 slice. This slice contains 9919 galaxies with brightness till isophotal magnitude $m_{iso} \leq$

*18.0*. 5127 galaxies from them have redshift measurement. If we adopt luminosity function given by Lin et al. (1996) [10] for generating random catalogue we obtain for $\xi(s)$ the power law with correlation length *6.25 h⁻¹Mpc* and $\gamma$=-*1.51* for the sample of galaxies with know redshift even in the case of first weight schemes (equal weight). It is in the very good agreement with results obtained by Tucker et al. (1997). It is very important, because it means that we could avoid complicated weight problem connected with the luminosity function. We should also notice that when we take into account all LCRS galaxies belonging to the -12 slice, we obtain slope's value -$\gamma$= -*1.41* and correlation length *5.5 h⁻¹Mpc* for galaxies with known redshift. It is also not far from values obtained by Tucker et al. (1997) [14].

Of course in the cases of the sample galaxies without known redshift, we must approximate the luminosity distance. We use the procedure explained in details by Machalski and Godłowski (2000) [15]. We should also notice that in Machalski and Godłowski (2000) [15] paper, a little different formula for luminosity function than Lin et al.(1996) [10] was obtained. When we generate a random catalogue with help of our luminosity function we obtain -$\gamma$=-*1.38* and correlation length *6 h⁻¹Mpc* for all galaxies and -$\gamma$=-*1.45* and correlation length *8.5 h⁻¹Mpc* for galaxies with known redshift. These results are presented in the fig.3. Now we start 3D analysis of the radiogalaxies. Our main result is that the correlation lengths are *3.75 h⁻¹Mpc* for the sample of galaxies with know redshift in all cases (all radio-identified galaxies, AGN and SB). However, slopes are different in all cases and we obtain the value -$\gamma$=-*1.76* for all galaxies (fig.4), -2.39 for AGN and -1.66 for SB.

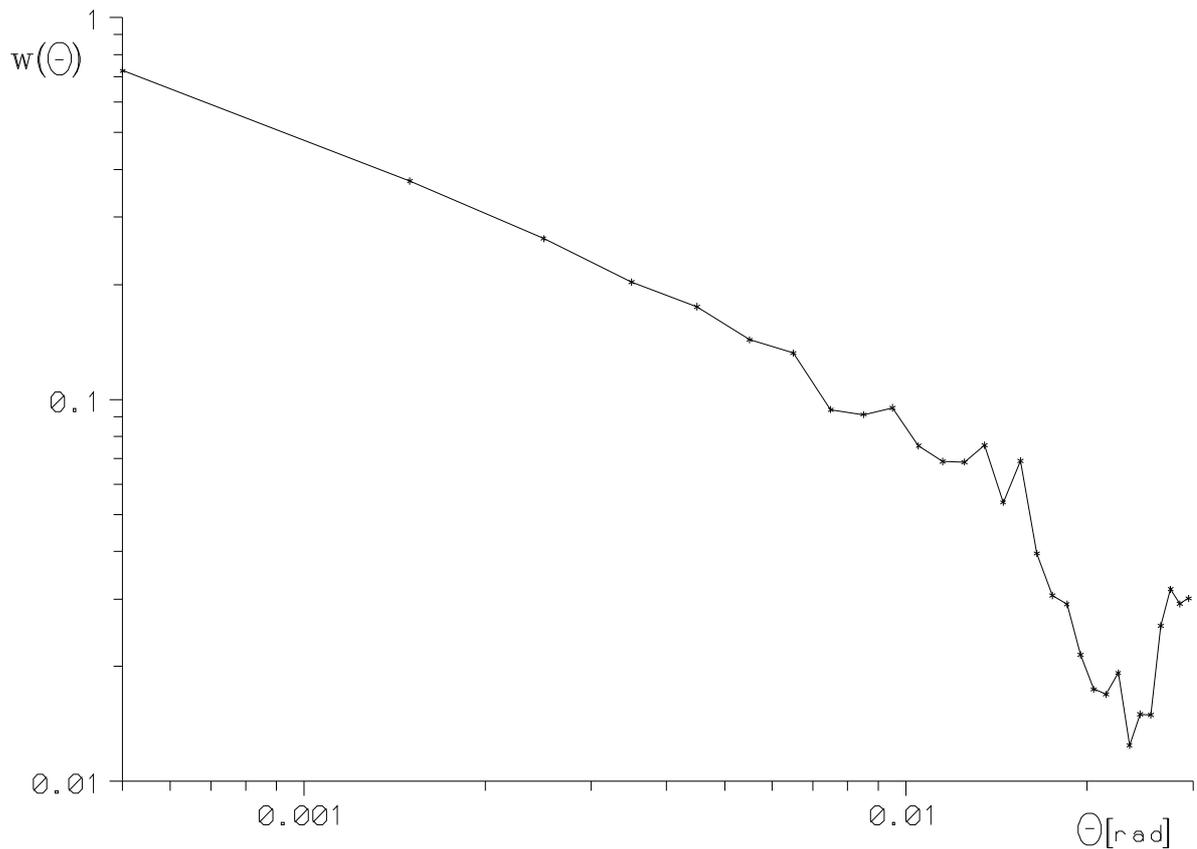

Fig.1 Angular correlation function for LSRS galaxies.

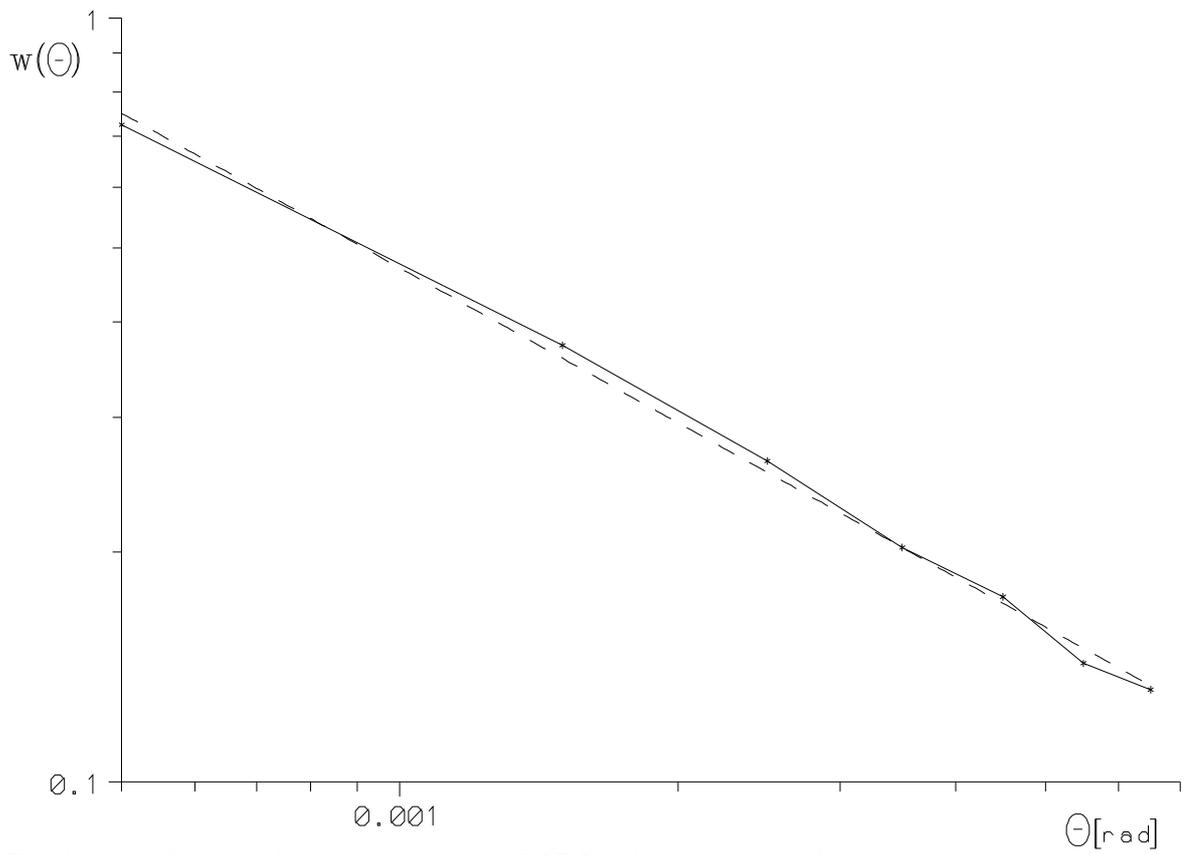
Fig.1a Angular correlation function for LSRS galaxies (power low area)

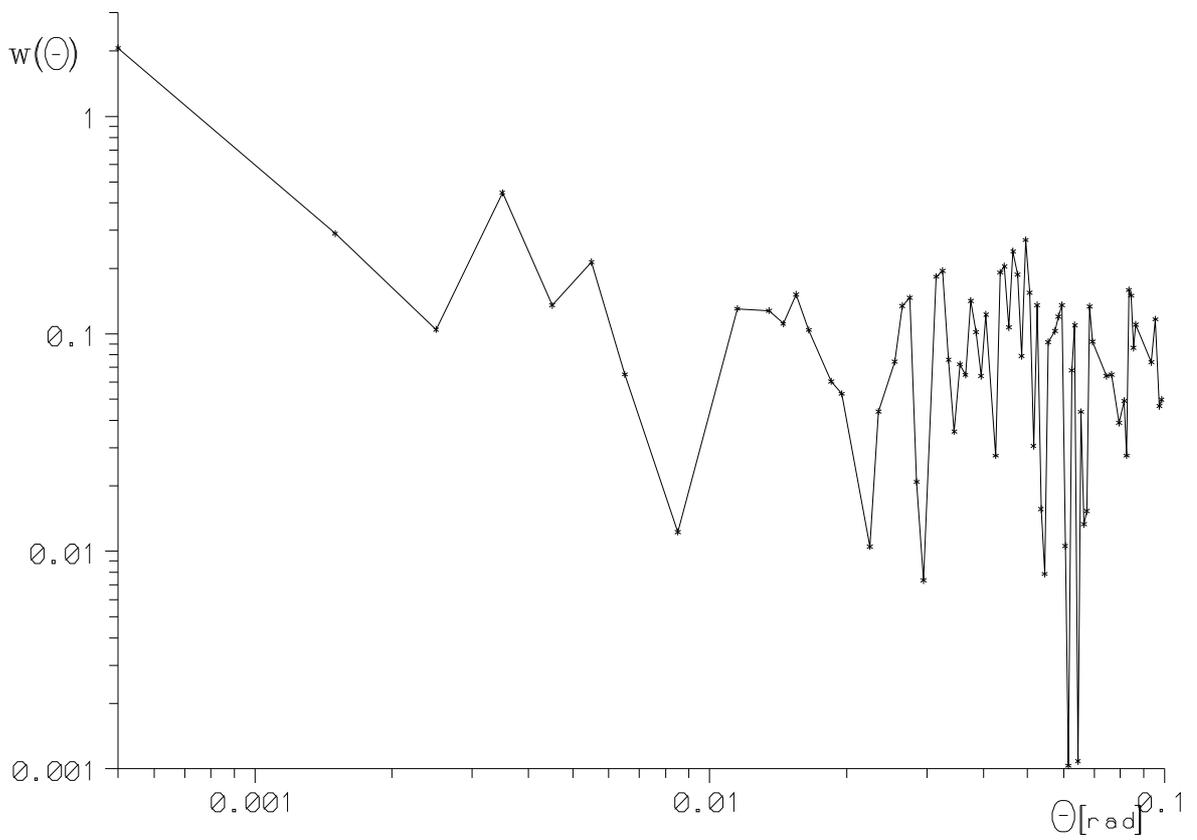
Fig.2 Angular correlation function for radio galaxies

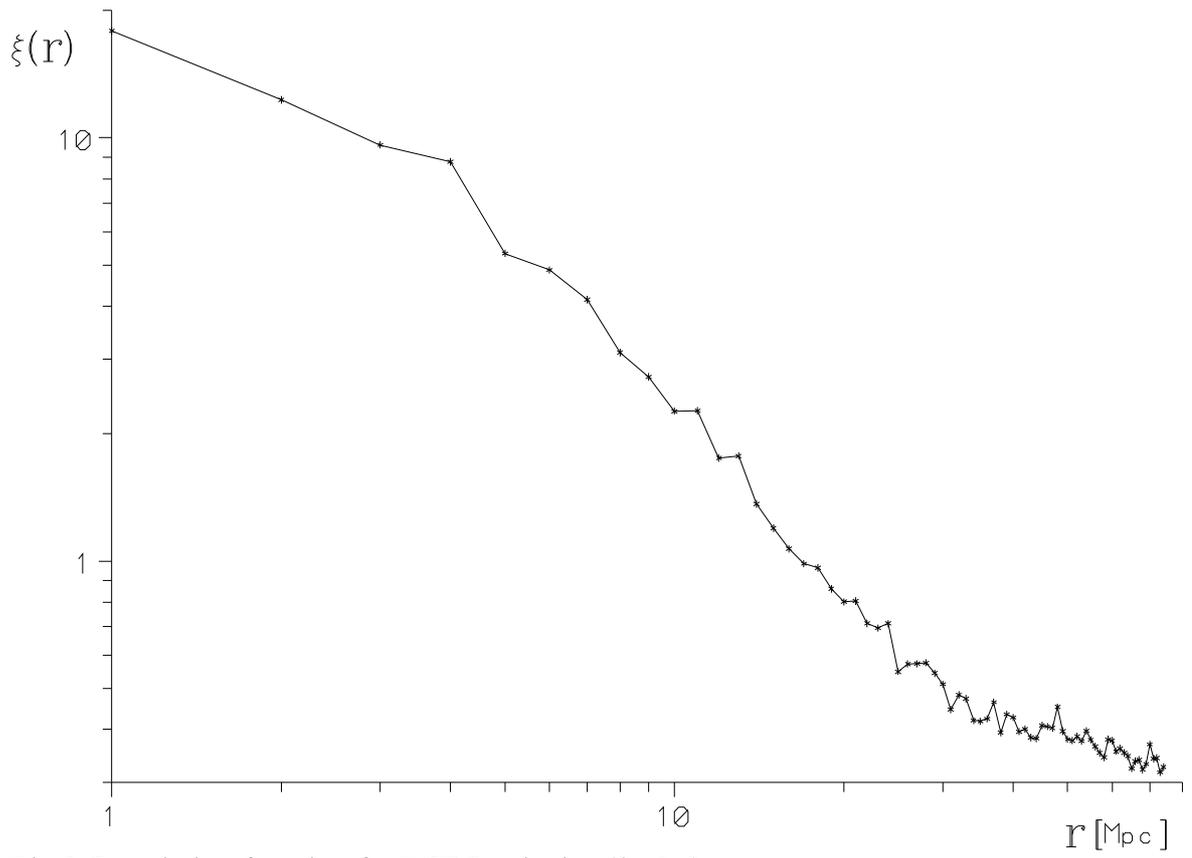

Fig.3 Correlation function for LCRS galaxies (*h=0.5*)

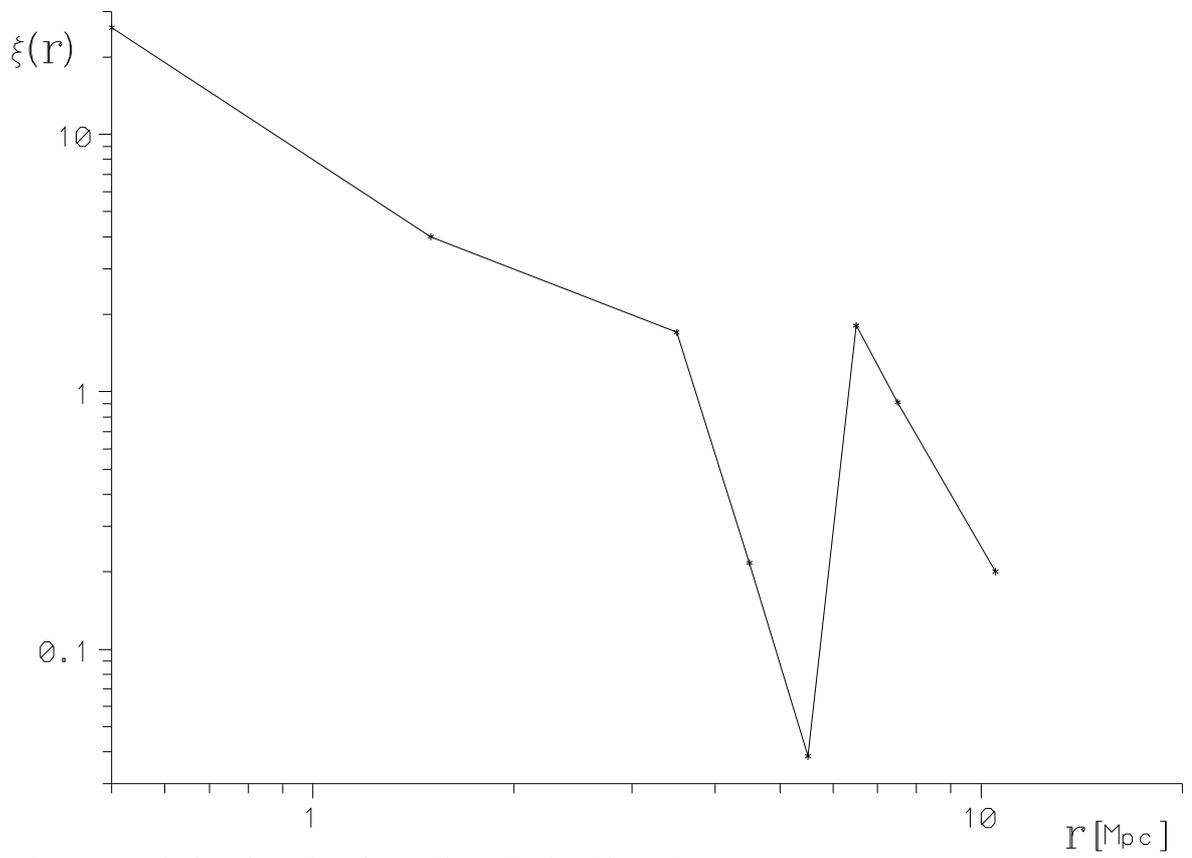

Fig.4 Correlation function for radio galaxies (*h=0.5*)

## 4. Conclusions

We have presented a detailed study of radiogalaxies clustering on the base of the sample of 1157 radio-identified galaxies (Machalski and Condon 1999) [1]. This sample was selected on the base of the Las Campanas Redshift Survey (LCRS) [2] and *1.4 Ghz NRA*)-VLA Sky Survey (NVSS) [3] catalog of the radio sources. Our main result is that for separation *2-15 $h^{-1}$ Mpc* autocorrelation function $\xi(s)$ can be approximated by the power law with slope *-$\gamma$=-1.8* and correlation length *3.75$h^{-1}$ Mpc*. This result means that correlation length for radiogalaxies is significantly smaller than found for whole sample of galaxies i.e. that radiogalaxies clusters more strongly than normal galaxies. The discovery of redshift clustering in our sample means that clustering of radiogalaxies is presented at least till *z~0.26* (deep of the LCRS).

## Acknowledgements

Author thanks prof J.Machalski for the permission of using his radiogalaxies data and the helpful discussion.